\documentclass[aps,reprint,nofootinbib,nobibnotes,notitlepage,superscriptaddress,twocolumn,prd,
preprintnumbers,
 amsmath,amssymb
]{revtex4-2}

\usepackage[caption=false]{subfig}
\usepackage{braket}
\usepackage{float}
\usepackage{lipsum}
\usepackage{graphicx}
\usepackage{dcolumn}
\usepackage{bm}
\usepackage{natbib}
\usepackage{hyperref}
\hypersetup{
	colorlinks = true,
    linkcolor = Red,
    urlcolor  = Red,
    citecolor = Red
}

\usepackage{microtype}

\usepackage{verbatim}
\usepackage[amssymb]{SIunits}
\usepackage{tabularx}
\usepackage[dvipsnames]{xcolor}
\usepackage{wasysym}

\usepackage{tikz}

\def\be{\begin{equation}}
\def\ee{\end{equation}}

\usepackage{color}
\definecolor{darkgreen}{RGB}{0,120,0}

\definecolor{darkgreen}{RGB}{0,120,0}

\newcommand{\resub}[1]{{#1}}

\newcommand{\hMpc}{h\,\mathrm{Mpc}^{-1}}

\def\beq{\begin{eqnarray}}
\def\eeq{\end{eqnarray}}

\makeatletter
\newlength{\apb@width}
\newcommand{\autoparbox}[2][c]{\settowidth{\apb@width}{#2}\parbox[#1]{\apb@width}{#2}}

\makeatother

\def\aap{\rm{A\&A}}

\def\jcap{\rm{JCAP}}

\begin{document}

\preprint{CERN-TH-2022-203}

\title{Constraining Single-Field Inflation with MegaMapper}

\author{Giovanni Cabass}
\email{gcabass@ias.edu}
\affiliation{School of Natural Sciences, Institute for Advanced Study, 1 Einstein Drive, Princeton, NJ 08540, USA}

\author{Mikhail M.~Ivanov}
\email{ivanov@ias.edu}
\affiliation{School of Natural Sciences, Institute for Advanced Study, 1 Einstein Drive, Princeton, NJ 08540, USA}
\affiliation{NASA Hubble Fellowship Program Einstein Postdoctoral Fellow}

\author{Oliver~H.\,E.~Philcox}
\email[Corresponding Author; ]{ohep2@cantab.ac.uk}
\affiliation{Center for Theoretical Physics, Department of Physics,
Columbia University, New York, NY 10027, USA}
\affiliation{Simons Society of Fellows, Simons Foundation, New York, NY 10010, USA}

\author{Marko Simonovi\'{c}}
\affiliation{Theoretical Physics Department, CERN,
1 Esplanade des Particules, Geneva 23, CH-1211, Switzerland}

\author{Matias Zaldarriaga}
\affiliation{School of Natural Sciences, Institute for Advanced Study, 1 Einstein Drive, Princeton, NJ 08540, USA}

\begin{abstract} 
    \noindent We forecast the constraints on single-field inflation from the bispectrum of future high-redshift surveys such as MegaMapper. Considering non-local primordial non-Gaussianity (NLPNG), we find that current methods will yield constraints of order $\sigma(f_{\rm NL}^{\rm eq})\approx 23$, $\sigma(f_{\rm NL}^{\rm orth})\approx 12$ in a joint power-spectrum and bispectrum analysis, varying both nuisance parameters and cosmology, including a conservative range of scales. Fixing cosmological parameters and quadratic bias parameter relations, the limits tighten significantly to $\sigma(f_{\rm NL}^{\rm eq})\approx 17$, $\sigma(f_{\rm NL}^{\rm orth})\approx 8$. These compare favorably with the forecasted bounds from CMB-S4: $\sigma(f_{\rm NL}^{\rm eq})\approx 21$, $\sigma(f_{\rm NL}^{\rm orth})\approx 9$, with a combined constraint of $\sigma(f_{\rm NL}^{\rm eq})\approx 14$, $\sigma(f_{\rm NL}^{\rm orth})\approx 7$; this weakens only slightly if one instead combines with data from the Simons Observatory. We additionally perform a range of Fisher analyses for the error, forecasting the dependence on nuisance parameter marginalization, scale cuts, and survey strategy. Lack of knowledge of bias and counterterm parameters is found to significantly limit the information content; this could be ameliorated by tight simulation-based priors on the nuisance parameters. The error-bars decrease significantly as the number of observed galaxies and survey depth is increased: as expected, deep dense surveys are the most constraining, though it will be difficult to reach $\sigma(f_{\rm NL})\approx 1$ with current methods. The NLPNG constraints will tighten further with improved theoretical models (incorporating higher-loop corrections \resub{and improved understanding of nuisance parameters}), as well as the inclusion of additional higher-order statistics.  
\end{abstract}

\maketitle

\paragraph*{Motivation}
The next generation of large-scale structure surveys will yield unprecedented measurements of the $z>2$ Universe. The combination of huge volumes and high-redshifts will enable proposed surveys, such as MegaMapper \citep{Schlegel:2019eqc,Schlegel:2022vrv}, MSE \citep{MSEScienceTeam:2019bva}, GAUSS \citep{Blanchard:2021ffq}, SpecTel \citep{Ellis:2019gnt,Dawson:2018fob} and Rubin \citep{Spergel:2015sza}, to constrain primordial physics by measuring a vast array of linear modes, improving on existing surveys by several orders of magnitude. A crucial question is the following: what can we hope to learn from this tranche of new data?

With precise data comes high-resolution measurements of the galaxy power spectrum, stretching to comparatively small (but yet still linear) scales. As with existing surveys, this can be used to place strong bounds on the cosmological model ($\nu\Lambda$CDM), via full-shape analyses \citep[e.g.,][]{2020JCAP...05..042I,2020JCAP...05..005D}, constraining parameters such as the matter density, primordial power spectrum amplitude, and Hubble constant. Such quantities are already relatively tightly constrained by the cosmic microwave background (CMB, \citep{2020A&A...641A...6P}), however, thus it is interesting to shift our attention to non-standard parameters, in particular those set by early Universe physics. As shown in \citep{Sailer:2021yzm}, by analyzing the high-redshift spectrum, we can hope to obtain strong constraints on particle physics, such as the mass of the neutrino and a number of proposed particles, such as axions, as well as energy deposition in the early Universe (via the $N_{\rm eff}$ parameter) \citep[e.g.,][]{Baumann:2011nk,Green:2021hjh,beutler19}. Furthermore, we can directly probe inflation by considering the spectral tilt, $n_s$, and its running, as well as the scale-dependent bias induced by local primordial non-Gaussianity ($f_{\rm NL}^{\rm loc}$), which is a key signature of multi-field inflation \citep[e.g.,][]{Desjacques:2008vf,Jeong:2009vd}.

By looking beyond the galaxy power spectrum, we can constrain a variety of other inflationary features. In particular, interactions in inflation and non-standard vacua can give rise to non-local primordial non-Gaussianity (NLPNG), whose primordial bispectra can be well described by the `equilateral' and `orthogonal' templates with amplitudes $f_{\rm NL}^{\rm eq}$ and $f_{\rm NL}^{\rm orth}$ \citep[e.g.,][]{Senatore:2009gt}. Careful analysis of the galaxy bispectra can yield constraints on these parameters, as demonstrated in \citep{Cabass:2022wjy,Cabass:2022ymb,DAmico:2022gki} for current data. Of course, the rich landscape of inflation is not limited to two templates: we may utilize the galaxy bispectrum to constrain a wealth of models, including massive spinning particles, and yet more can be learnt from the galaxy trispectrum \citep[e.g.,][]{2015arXiv150308043A,MoradinezhadDizgah:2018ssw,Cabass:2022oap}. For now, the constraining power on such parameters is dominated by the CMB \citep[e.g.,][]{2020A&A...641A..10P,2020A&A...641A...9P}; however, with the advent of stage-five spectrosopic surveys, the attention will shift to large scale structure (LSS). In this work, we forecast how well proposed surveys such as MegaMapper can hope to constrain inflationary signals, via the NLPNG parameters, optionally in conjunction with CMB observations.

\vskip 8pt
\paragraph*{Set-Up}
To forecast the efficacy of high-redshift in constraining single-field inflationary parameters, we follow a similar procedure to \citep{Sailer:2021yzm}, concentrating on the proposed MegaMapper experiment outlined in \citep{Ferraro:2019uce}. Unlike previous work, we include a full treatment of the bispectrum, modeled in conjunction with the power spectrum to minimize parameter degeneracies. The fiducial experiment includes $\approx 40$ million galaxies in a redshift range $2<z<5$, which we divide into four contiguous bins with number density and redshift defined by Tab.~1 of \citep{Ferraro:2019uce}. In each redshift bin, we compute a fiducial power spectrum and bispectrum using the parameters of \citep{Sailer:2021yzm}, in particular tidal biases set by the coevolution model \citep{Desjacques:2016bnm} and an (optimistic) fingers-of-God (FoG) dispersion $\sigma_{v}=100\,\mathrm{km}\,\mathrm{s}^{-1}$. For the quadratic bias, we use the fitting formula of \citep{Lazeyras:2015lgp}, which is significantly more accurate than the coevolution prediction for highly biased samples.

We utilize the one-loop power spectrum and the tree-level bispectrum model summarized in \citep{Ivanov:2019pdj,Chudaykin:2020ghx,Ivanov:2021kcd}, depending on the following nuisance parameters:
\beq\label{eq: bias-params}
	\{b_1,b_2,b_{\mathcal{G}_2},b_{\Gamma_3},c_0,c_2,c_4,\tilde{c},c_1,P_{\rm shot},B_{\rm shot},a_0,a_2,b_\phi\}
\eeq
describing linear, quadratic and tidal bias, five counterterms, four stochasticity parameters, and non-Gaussian bias. These are subject to wide Gaussian priors following \citep{2021arXiv211204515P,Cabass:2022wjy}. \resub{We do not include one-loop corrections to the bispectrum; whilst these are available for $\Lambda$CDM \citep{Philcox:2022frc,DAmico:2022osl}, the loops involving $f_{\rm NL}$ have yet to be self-consistently included.} We include power spectrum multipoles up to $\ell=4$ and the bispectrum monopole, noting that constraints may tighten somewhat if we include higher-order bispectrum multipoles \citep{Ivanov:2023qzb}. We additionally include the $Q_0$ statistic of \citep{Ivanov:2021fbu} (see also \citep{Scoccimarro:2004tg,DAmico:2021ymi}), which is a proxy for the real-space power spectrum, and allows extraction of information beyond the usual fingers-of-God limits. We vary the following cosmological parameters:
\beq\label{eq: cosm-params}
	\{h, \omega_{\rm cdm}, \log 10^{10}A_s, f_{\rm NL}^{\rm eq}, f_{\rm NL}^{\rm orth}\}, 
\eeq
with wide flat priors, where the last two parameters control the single-field inflation model. We do not include the multi-field parameter, $f_{\rm NL}^{\rm loc}$, since the information content on this is dominated by the power spectrum \citep{Cabass:2022ymb} and is not strongly degenerate with the NLPNG amplitudes. We additionally fix the spectral tilt, $n_s$, \resub{since it will be precisely constrained by future CMB surveys. If one instead marginalizes over $n_s$, the $f_{\rm NL}$ constraints degrade by $<5\%$, as verified in MCMC analyses (similar to the conclusions of \citep{Cabass:2022wjy} for the BOSS survey)}.

\begin{table}
    \centering
    \begin{tabular}{l||c|c}
    \hline
    \textbf{Experiment} & $\sigma(f_{\rm NL}^{\rm eq})$ & $\sigma(f_{\rm NL}^{\rm orth})$\\ \hline\hline
    MegaMapper - A   & 23 & 10 \\
    MegaMapper - B   & 22 & 10 \\
    MegaMapper - C   & 17  & 8 \\\hline
    Planck 2018  &  47 & 24\\
    Simons Observatory (SO) & 27 & 14\\
    CMB-S4 & 21 & 9\\\hline
    MegaMapper + SO & 16 & 8\\
    MegaMapper + CMB-S4 & 14 & 7\\\hline
    \end{tabular}
    \caption{$68\%$ Constraints on the non-local primordial non-Gaussianity amplitudes shown in Fig.\,\ref{fig: fNL1}\,\&\,\ref{fig: fNL2}. The three MegaMapper analyses are (A) free cosmology and bias parameters, (B) fixed cosmology and free bias parameters, (C) fixed cosmology and fixed quadratic bias relations. We additionally quote the \textit{Planck} 2018 constraints, as well as the forecasts for the Simons Observatory and CMB-S4. The final entry gives the joint constraints from MegaMapper and future CMB experiments, assuming that the latter datasets fix the cosmological parameters.} 
    \label{tab: fNL}
\end{table}

Following \citep{Sailer:2021yzm}, we fix the minimum power spectrum wavenumber to $k_{\rm min}=\mathrm{max}[0.003\hMpc,2\pi/V^{1/3}]$, and use bins of width $\Delta k = 0.005\hMpc$. We fix the maximum scale by asserting that the relative size of the FoG term (which usually dominates the theoretical error) is the same as for the BOSS analysis with $k^P_{\rm max}\approx 0.17\hMpc$ \citep{Cabass:2022wjy}. This is a relatively conservative choice and is equivalent to demanding that the FoG contributions are at most $10\%$ of the tree-level theory. The $Q_0$ statistic is not affected by FoG, thus we consider modes up to the Zel'dovich velocity dispersion scale, fixing $k_{\rm max}^Q  = \left[\int dq P_{\rm lin}(q)/(6\pi^2)\right]^{-1/2}$, again following \citep{Sailer:2021yzm}.\footnote{This is a conservative choice, since $\left[\int dq P_{\rm lin}(q)/(6\pi^2)\right]^{-1/2}$ in question is the characteristic scale of the dispacement field, the bulk of which is accounted for via infrared resummation \citep{Blas:2016sfa,Ivanov:2018gjr,Senatore:2014via}. The true $k_{\rm max}$ relevant for $Q_0$ can be significantly larger than this (up to the non-linear scale) as discussed in \citep[e.g.,][]{Ivanov:2021fbu}. Since $Q_0$ does not contribute strongly to NLPNG constraints, this choice is not strongly relevant for our analysis.} For the bispectrum, we consider broader bins of $\Delta k = 0.01\hMpc$ (since the desired signal is relatively smooth), and fix $k_{\rm min}^B=0.01\hMpc$, and the maximum by asserting that the FoG contribution has similar contributions to those in the BOSS analyis, which here requires it to be less than $2\%$ of the tree-level theory, yielding $k_{\rm max}^B\approx 0.20\hMpc$ at $z=3$. This is lower than for the power spectrum, since we include only tree-level terms in the bispectrum model (though could extend beyond this via the approaches of \citep{Philcox:2022frc,DAmico:2022osl}).\footnote{Imposing stronger bounds on the FoG contribution reduces the detection significance somewhat: we find a $\approx30\%$ increase in $\sigma(f_{\rm NL})$ when restricting the terms to be $5\%$ and $1\%$ of the power spectrum and bispectrum respectively. This is equivalent to increasing $\sigma_v$ by a factor of $\sqrt{2}$. Further discussion of the dependence on $k_{\rm max}$ can be found below.} Finally, we perform the forecast via an MCMC analysis of the fiducial $f_{\rm NL}^{\rm eq}=f_{\rm NL}^{\rm orth}=0$ spectra, using the \textsc{Class-PT} code \citep{2020PhRvD.102f3533C}, the \textsc{MontePython} sampler \citep{Brinckmann:2018cvx} and the public likelihoods described in \citep{2021arXiv211204515P},\footnote{Available at \href{https://github.com/oliverphilcox/full_shape_likelihoods}{github.com/oliverphilcox/full\_shape\_likelihoods}.} assuming a Gaussian likelihood. 

\resub{In this \textit{Letter}, we assume a diagonal covariance matrix for the power spectrum multipoles, as in \citep{2019JCAP...11..034C}. In principle, late-time non-Gaussianity leads to correlations both within and between correlators \citep{Sugiyama:2019ike}, which could somewhat reduce the detection significances (as shown in line-intensity-mapping contexts in \citep{Floss:2022wkq}). For spectroscopic surveys, the full impacts of these on $f_{\rm NL}$ constraints have yet to be explored; however, given our conservative choices of $k_{\rm max}$ and the high shot-noise of MegaMapper-like surveys, we expect this to be small, given the lack of importance of non-Gaussian covariance in power spectrum \citep{2020PhRvD.102l3521W} and large-scale bispectrum \citep{Ivanov:2021kcd,Philcox:2022frc,2021arXiv211204515P} $\Lambda$CDM analyses.}

\begin{figure}
    \centering
    \includegraphics[width=0.45\textwidth]{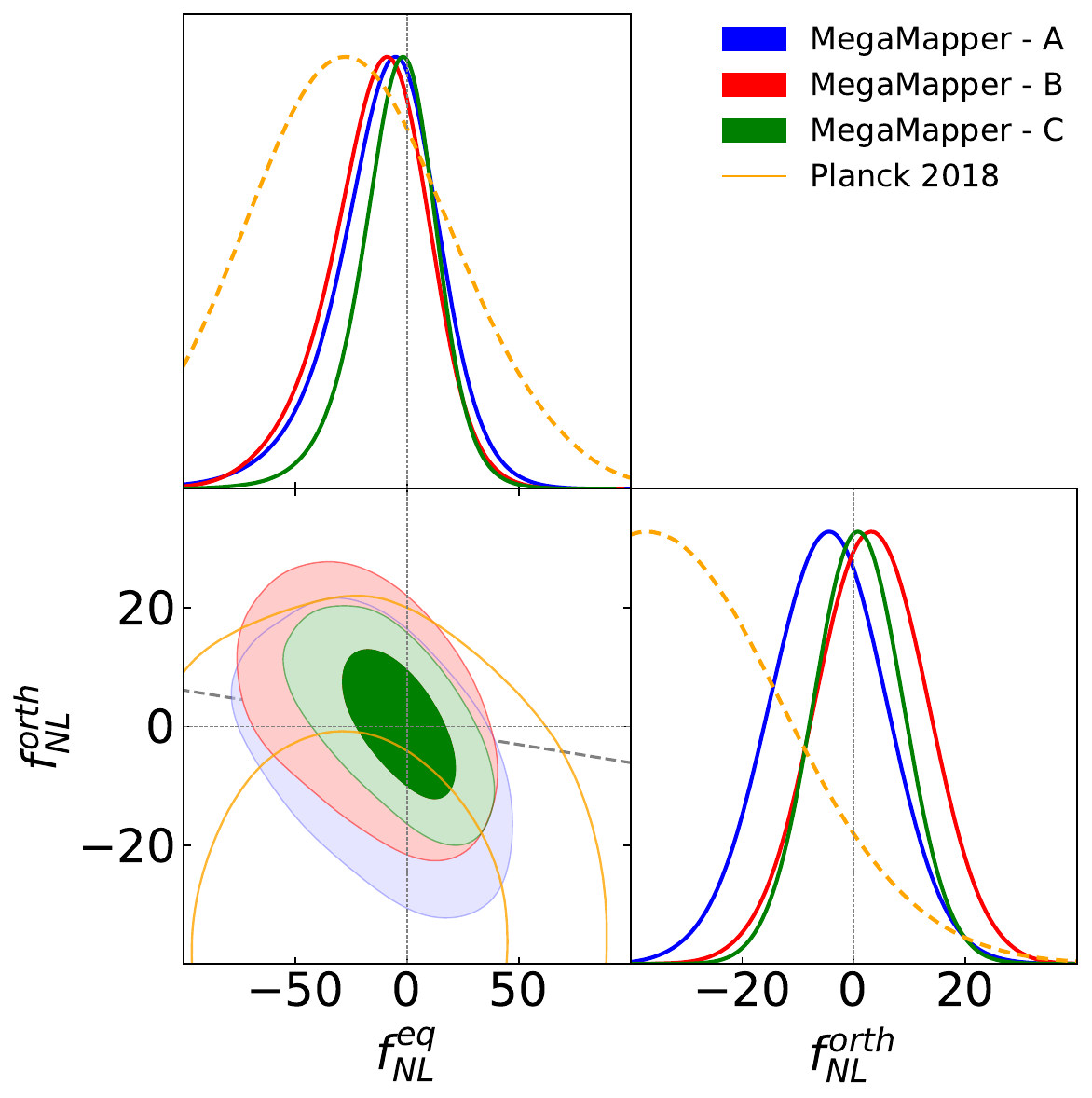}
    \caption{Forecasted constraints on the non-local primordial non-Gaussianity amplitudes, $f_{\rm NL}^{\rm eq}$, $f_{\rm NL}^{\rm orth}$ from the proposed MegaMapper high-redshift experiment \citep{Schlegel:2022vrv}, alongside measurements from \textit{Planck} \citep{2020A&A...641A...9P}. The MegaMapper forecasts are computed via MCMC using the parameters described in the text, including both the one-loop power spectrum and tree-level bispectrum. We consider three types of forecast: A (blue), varying both cosmological and nuisance parameters; B (red) fixing cosmological parameters; C (green) fixing also quadratic bias relations. Numerical constraints are given in Tab.\,\ref{tab: fNL}, and we show the linear relationship between $f_{\rm NL}^{\rm orth}$ and $f_{\rm NL}^{\rm eq}$ relevant for DBI inflation as a dashed line \citep{Alishahiha:2004eh,Cabass:2022wjy}.}
    \label{fig: fNL1}
\end{figure}

\vskip 8pt
\paragraph*{Fiducial Constraints}
We consider three characteristic cases for the inflationary forecasts, similar to \citep{Cabass:2022wjy}. Firstly, we assume no knowledge of cosmology or bias, varying all parameters in \eqref{eq: bias-params}\,\&\,\eqref{eq: cosm-params}. Secondly, we assume that the background $\nu\Lambda$CDM cosmology is known (for example from the power spectrum, or from external data e.g., CMB-S4), and vary only biases. Thirdly, we additionally fix the quadratic bias parameter relations ($b_2(b_1)$ and $b_{\mathcal{G}_2}(b_1)$); these generate shapes with significant correlations with NLPNG \citep[e.g.,][]{Baumann:2021ykm}, and could be potentially fixed via tight simulation-derived priors.\footnote{One may also place priors on the non-Gaussian biases, such as $b_\phi$: this has limited effect for NLPNG (but is crucial for $f_{\rm NL}^{\rm loc}$, see \citep{Barreira:2020ekm,Barreira:2022sey}), as seen by the similar constraints from MCMC and Fisher forecasts in the below, noting that the latter require fixed $b_\phi$ (as it appears only proportional to $f_{\rm NL}$). This is consistent with the forecast of \citep{Gleyzes:2016tdh}, which found $\sigma(f_{\rm NL}^{\rm equil})\approx 500$ from the power spectrum scale-dependent bias alone, for a MegaMapper-type survey.}

\begin{figure}
    \centering
    \includegraphics[width=0.45\textwidth]{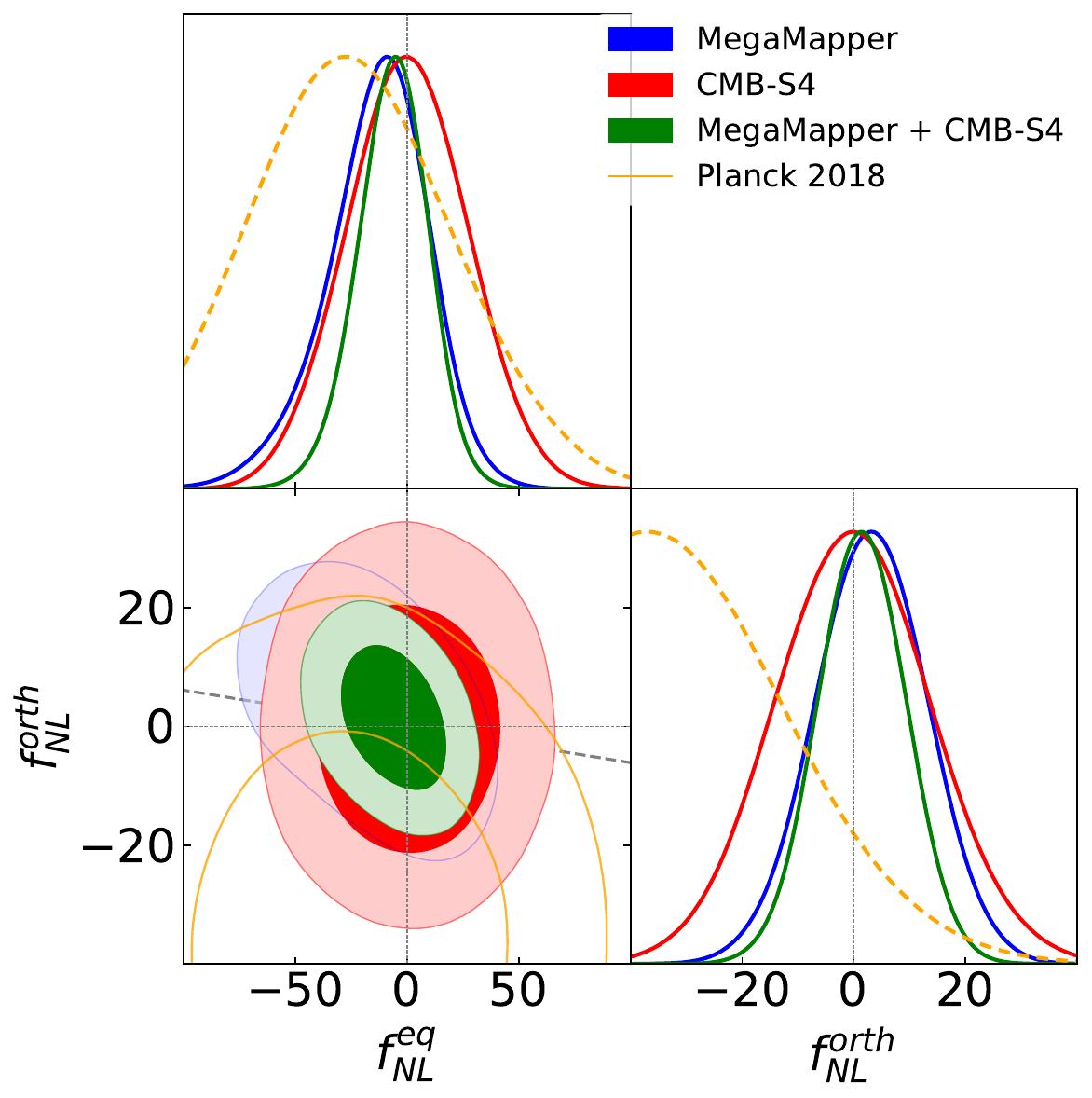}
    \caption{Comparison of MegaMapper (blue) constraints on $f_{\rm NL}^{\rm eq}$ and $f_{\rm NL}^{\rm orth}$ with those from CMB-S4 \citep{CMB-S4:2016ple} (red). The joint constraint (green) is obtained by combining the fixed cosmology MegaMapper contour with the CMB-S4 forecasted posterior.}
    \label{fig: fNL2}
\end{figure}

The corresponding constraints on $f_{\rm NL}$ are given in Fig.\,\ref{fig: fNL1} and Tab.\,\ref{tab: fNL}. In the conservative case, we find $\sigma(f_{\rm NL}^{\rm eq}) \approx 23$ and $\sigma(f_{\rm NL}^{\rm orth})\approx 10$; neither constraints are found to tighten significantly when we fix the fiducial cosmology (and thus the NLPNG templates), but we see a modest ($\approx 30\%$) improvement by fixing the quadratic bias relations, and thus reducing the relevant degeneracies. In the most optimistic case, we find $\sigma(f_{\rm NL}^{\rm eq})\approx 17$, $\sigma(f_{\rm NL}^{\rm orth}) \approx 8$, though we note that the relevant bias parameter relation priors may be difficult to obtain in practice (since the data already constrain the tidal bias to $\approx 10\%$, for example). Note that the impact of bias parameter priors is weaker here than in \citep{Cabass:2022wjy} due to the larger $k$ range, and thus greater internal degeneracy breaking.

To place our results in context, we may compare them to current limits from the CMB and LSS. The BOSS data constrain $f_{\rm NL}^{\rm eq} = 260\pm300, f_{\rm NL}^{\rm orth}=-23\pm120$ \citep{Cabass:2022wjy} (with fixed biases), which are considerably weaker than the CMB bounds of $f^{\rm eq}_{\rm NL} = -26\pm 47, f^{\rm orth}_{\rm NL} = -38\pm 24$ from \textit{Planck} 2018 \citep{2020A&A...641A...9P}. The next generation of CMB data is expected to tighten these bounds somewhat (cf.\,Fig.\,\ref{fig: fNL2}), with $\sigma(f^{\rm eq}_{\rm NL}) = 27, \sigma(f^{\rm orth}_{\rm NL}) = 14$ expected from the Simons Observatory (SO) \citep{2019JCAP...02..056A}, and $\sigma(f^{\rm eq}_{\rm NL}) = 21, \sigma(f^{\rm orth}_{\rm NL}) = 9$ from CMB-S4 \citep{Abazajian:2019tiv,CMB-S4:2022ght,CMB-S4:2016ple}. This improvement is relatively small given that \textit{Planck} is already cosmic variance limited for the large-scale temperature modes, and there is little gain from small scales due to the numerous secondary contributions.  

The forecasts above improve upon LSS constraints by around an order of magnitude, and, even in the conservative case, where we vary cosmology and all bias parameters, are significantly tighter than those of \textit{Planck}. This matches expectations, since the survey volume of MegaMapper is roughly $30\times$ that of BOSS, and we work at higher redshift, facilitating larger $k_{\rm max}$. If the $\nu\Lambda$CDM parameters are known (or at least highly constrained), MegaMapper can provide competitive constraints to CMB-S4, and somewhat tighter than the SO baseline. In combination, the two yield stronger constraints still, as shown in Fig.\,\ref{fig: fNL2}, with bounds of $\sigma(f_{\rm NL}^{\rm eq})\approx 14$, $\sigma(f_{\rm NL})\approx 7$ expected from CMB-S4, broadening to $\sigma(f_{\rm NL}^{\rm eq})\approx 16$, $\sigma(f_{\rm NL})\approx 8$ with SO. We may also compare the results to previous simplified forecasts, in particular \citep{Ferraro:2019uce}, based on \citep{Karagiannis:2018jdt}. This obtained $\sigma(f_{\rm NL}^{\rm eq}) = 40$, $\sigma(f_{\rm NL}^{\rm orth})=9$, varying all relevant parameters: these agree with our fiducial analysis (`A') to $\approx 40\%$; a factor certainly appropriate for Fisher forecasts. Of course, these results depend on the various modeling choices shown above, in particular the $k$ ranges and fiducial bias and FoG parameters. However, our wavenumber limits may be regarded as conservative, and the modelling can likely be improved further by the addition of one-loop terms, \resub{though it is uncertain whether this will significantly impact parameter constraints \cite{Philcox:2022frc}.}

\vskip 8pt
\paragraph*{Impact of Analysis Choices}
To better understand the above results, it is useful to consider the dependence of the NLPNG constraints on the maximum wavenumber used in the analysis, $k_{\rm max}^{P,B}$, and the assumed priors on nuisance parameters. To this end, we perform a Fisher analysis based on the above methodology, forecasting $\sigma(f_{\rm NL}^{\rm eq})$ and $\sigma(f_{\rm NL}^{\rm orth})$ for the fiducial MegaMapper survey, using the same galaxy distribution parameters parameters as before. For the fiducial MegaMapper set-up, the Fisher forecast matches the full MCMC constraints to $\approx 10\%$, at significantly lower computational cost. 

Fisher analyses are performed for nine values of $k_{\rm max}^B\in[0.05,0.5]\hMpc$, using the same scale-cut for each redshift bin and fixing $k_{\rm max}^P = 2k_{\rm max}^B$, motivated by the above physical limits. In each case, we consider two scenarios: one in which all nuisance parameters (encompassing bias, stochasticity, and counterterms) are varied, akin to method `B' in the MCMC forecasts, and one in which only $f_{\rm NL}^{\rm eq}$ and $f_{\rm NL}^{\rm orth}$ are free, with all other parameters fixed to their fiducial values. Whilst the extent to which future analyses can strongly bound nuisance parameters is unclear, this provides a practical bound to the precision of NLPNG measurements, in the limit of perfect understanding of the UV physics of galaxy formation, as well as nonlinear gravitational collapse.

\begin{figure}
    \centering
    \includegraphics[width=0.5\textwidth]{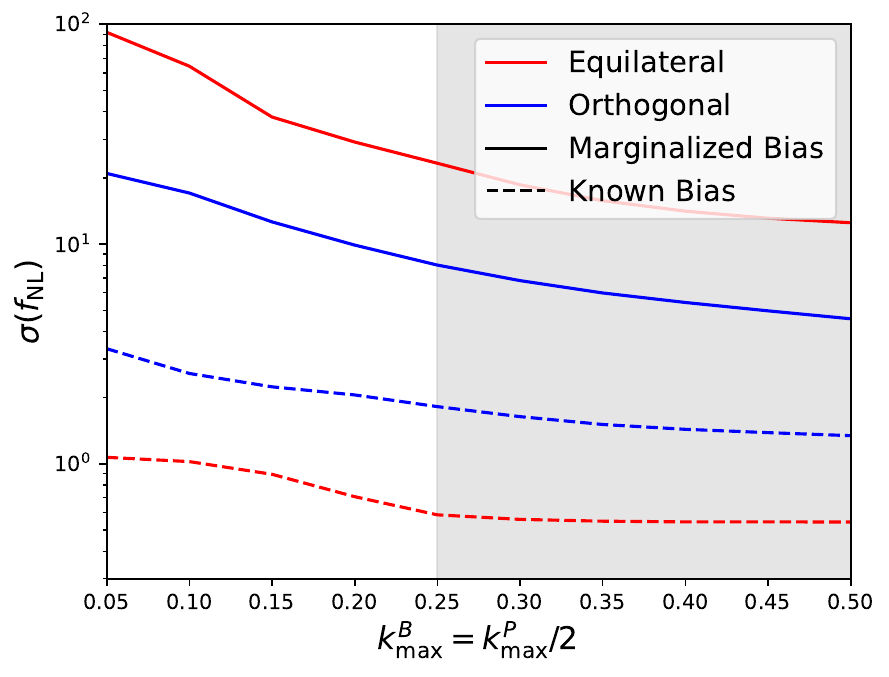}
    \caption{Impact of scale cuts and nuisance parameter marginalization on the $f_{\rm NL}$ constraints obtained from an idealized MegaMapper-like survey. Fisher forecasts are shown for both equilateral (red) and orthogonal (blue) primordial non-Gaussianity, marginalizing over all NLPNG and nuisance parameters (full lines) or only the NLPNG parameters (dashed lines), jointly varying both the maximum power spectrum and bispectrum wavenumber across all redshift slices. We show the fiducial MegaMapper constraints as horizontal dotted lines (cf.\,Tab.\,\ref{tab: fNL}), noting that these use a different $k_{\rm max}$ for each redshift bin, and show in grey the rough range of scales for which higher-loop effects become important. The plot indicates that $f_{\rm NL}$ constraints are strongly limited by our lack of knowledge of nuisance parameters, and show considerable dependence on $k_{\rm max}$. Whilst the known-bias case represents an optimistic lower bound on the parameter error, we caution that higher-loop terms (both in hydrodynamic and gravitational physics) will be required to extend the modelling to large $k_{\rm max}$, and that precise knowledge of nuisance parameters will require accurate hydrodynamic simulations coupled with good understanding of observational effects.}
    \label{fig: kmax-plot}
\end{figure}

The results are shown in Fig.\,\ref{fig: kmax-plot}, and indicate considerable dependence of the error-bar on $k_{\rm max}$. This is unsurprising, since the signal-to-noise of the statistics increases considerably with $k$ (up to the shot-noise dominated regime), allowing breaking of parameter degeneracies. At large $k_{\rm max}$, we caution that additional loops (and thus nuisance parameters) will be required for accurate modelling, which will somewhat temper the reduction in $\sigma(f_{\rm NL})$; this idealized plot indicates the utility of extending the modelling further however. We also find a considerable improvement in constraining power (at least an order of magnitude) when the nuisance parameters are known, particularly for equilateral non-Gaussianity. This implies that the majority of the signal-to-noise goes into constraining parameters such as higher-order bias and shot-noise rather than NLPNG directly (particularly at low $k_{\rm max}$, whereupon information from the power spectrum dominates), and motivates further study into simulation-based priors on such effects. If one had perfect knowledge of the nuisance parameters and non-linear physics (which will likely be challenging, even in the far future), constraints of $\sigma(f_{\rm NL})=\mathcal{O}(1)$ could be obtained from a MegaMapper-like survey. 

\vskip 8pt
\paragraph*{Survey Design}
Finally, it is interesting to consider what types of survey could yield the best constraints on NLPNG parameters. This can also be probed using Fisher forecasts, here considering an idealized Lyman-Break galaxy survey (LBG, modelled using the $n(z)$ of \citep{Ferraro:2019uce}), and computing $\sigma(f_{\rm NL})$ as a function of the total number of observed galaxies, $N_g$, and the maximum redshift, $z_{\rm max}$ (using four redshift bins, as before). 

\begin{figure*}
    \centering
    \includegraphics[width=0.9\textwidth]{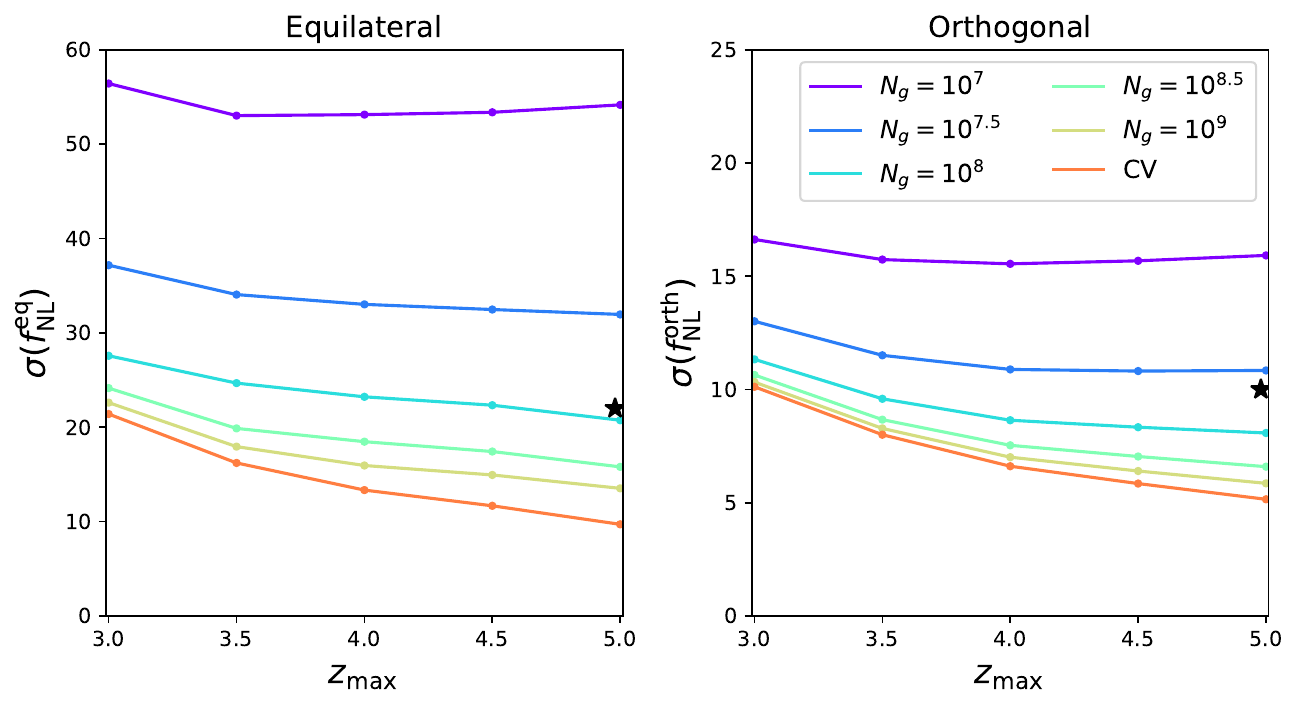}
    \caption{Fisher forecast of the error-bar on NLPNG parameters as a function of the maximum survey redshift ($z_{\rm max}$) and the total number of galaxies ($N_g$). 30 results are shown, with the density indicated by the caption, where `CV' indicates an idealized cosmic-variance limited sample. The MCMC forecasts appropriate for a MegaMapper-like experiment (Tab.\,\ref{tab: fNL}) are shown as black stars. In each case, we fix cosmology, but allow all bias and nuisance parameters to vary freely, in four redshift bins. Increasing the survey volume at fixed $N_g$ yields a slight increase in precision, with a much larger one seen by increasing the number of target galaxies.}
    \label{fig: highz}
\end{figure*}

The results are shown in Fig.\,\ref{fig: highz}, displaying 30 (fixed-cosmology) forecasts, with $z_{\rm max}\in[3,5]$, $\log_{10} N_g\in [7,9]$, and an additional cosmic-variance dominated sample (with the idealized limit of $N_g\to\infty)$.\footnote{In practice, there are only a finite number of LBGs in the Universe, thus $\log_{10} N_g\lesssim8$. Here, we allow arbitrarily large $N_g$ to emulate other galaxy populations one may wish to study.} At small $N_g$, we find little gain in pushing to large volumes due to the prohibitively low sample density; however, as $N_g$ increases, we see significant reduction in $\sigma(f_{\rm NL})$ as the survey volume increases for a fixed number of targets. This occurs due to the trade-off between low cosmic variance (scaling as the root of the survey volume, $V^{1/2}$) and low shot-noise (depending on $N_g/V$). In contrast, increasing the total number of galaxies observed (at fixed $V$) yields much increased sensitivity at all $z_{\rm max}$, with the cosmic variance limit yielding $\sigma(f_{\rm NL}^{\rm eq})\approx 10$, $\sigma(f_{\rm NL}^{\rm orth})\approx 5$. Even with futuristic surveys such as MegaMapper, it seems that we are far from this limit, and are thus not shot-noise dominated on all scales of interest; approaching  would require the inclusion of additional galaxy samples however.

\vskip 8  pt
All of our forecasts indicate that future high-redshift surveys will yield strong constraints on NLPNG. These are likely to surpass those from the CMB, and are made possible by the analysis of higher-order statistics. This is particularly true for the non-local non-Gaussianity parameters: whilst similar forecasts to the above can be performed for $f_{\rm NL}^{\rm loc}$, we expect that the bulk of the signal-to-noise will come from the power spectrum, with the bispectrum adding only $\approx 30\%$ \citep{Cabass:2022ymb}. Our results, however, have strong dependence on the $k$ ranges assumed, which themselves depend on (poorly understood) MegaMapper sample parameters. To obtain a better understanding of the capabilities of future surveys, it will be vital to generate accurate mock catalogs appropriate for the high-redshift regime. We close by stressing that the three $f_{\rm NL}$ parameters do not paint a full picture of inflation: MegaMapper, or some similar experiment, will allow a wide variety of models to be probed, including ghosts, colliders, and axions.

\vskip 8pt
\paragraph*{Acknowledgements}
We thank Daniel Green and the UCSD ``Primordial Physics with Spectroscopic Surveys'' workshop for encouraging us to write this forecast. \resub{We additionally thank the anonymous referee for insightful feedback.} OHEP is a Junior Fellow of the Simons Society of Fellows.
The work of MMI has been supported by NASA through the NASA Hubble Fellowship grant \#HST-HF2-51483.001-A awarded by the Space Telescope Science Institute, which is operated by the Association of Universities for Research in Astronomy, Incorporated, under NASA contract NAS5-26555. GC acknowledges support from the Institute for Advanced Study. 

\appendix

\bibliographystyle{apsrev4-1}
\bibliography{refs}

\end{document}